# SVRPF: An Improved Particle Filter for a Nonlinear/non-Gaussian Environment

**XINGZI QIANG, YANBO ZHU, AND RUI XUE*, (Member, IEEE)**

Beihang University, Beijing 100191, China

*Corresponding author: Rui Xue (e-mail: xuerui@ buaa.edu.cn).

This work was supported in part by the National Key Research and Development Program of China under Grant 2017YFB0503401 and in part by the National Natural Science Foundation of China under Grant U1833125 and Grant 61803037.

**ABSTRACT** The performance of a particle filter (PF) in nonlinear and non-Gaussian environments is often affected by particle degeneracy and impoverishment problems. In this paper, these two problems are re-assessed using the concepts of importance region (***IR***) selection and particle density (*PD*), where ***IR*** describes the distribution region of particles, and *PD* describes the density of particles in ***IR***. Based on these two factors, a support vector regression PF (SVRPF) is proposed to overcome the problems from nonlinear and non-Gaussian environments, especially in regard to narrow observation noise. Furthermore, the consistency of the SVRPF and Bayes' filtering is demonstrated. A numerical simulation shows that the performance of the SVRPF is more stable than other filter algorithms. Provided that other conditions are the same, when the observation noise variance is 0.1 and 5, the root-mean-square errors of the SVRPF decrease by 0.5 and 0.03, respectively, compared with that of a general PF.

## I. INTRODUCTION

The particle filter (PF), which is used to estimate the current state of a system, has been demonstrated to be a potential technique for nonlinear and non-Gaussian environments [1–4]. In the last few years, the PF and its improvements have been widely used to solve different nonlinear and non-Gaussian Bayesian filtering problems arising in different areas, such as target detection [5,6], visual tracking [7], fault detection [8], channel estimation [9], melody extraction [10], life prediction [11], among others [12,13], primarily because of its capability to approximate any probability density function (PDF). Nevertheless, the particle degeneracy and impoverishment problem are two serious drawbacks of the PF that has aroused a great deal of interest from many researchers [14–17].

References [18] and [19] introduce respectively EKF [20] and UKF [21] to the PF framework. The core idea of these two methods is to optimize the proposal distribution to make them closer to the real posterior-PDF. To a large extent, these two improved PF algorithms achieve good results in applications, gaining wide-spread use in practical engineering [22,23]. References [24] and [25] propose the iterated extended Kalman filter and the iterated unscented Kalman filter to improve the proposal distribution, and approximate the real posterior-PDF. The feedback PF was designed based on an ensemble of a controlled stochastic system (the particles) [26]. In this method, the evolution of each particle is controlled by its own state and feedback circuit. In [4], the PF was improved by introducing the genetic algorithm to overcome the impoverishment problem. All the above methods are geared to optimize the proposal distribution based on introducing the current observation.

In addition, because of the advantage of the resampling technique in solving the degeneracy problem, various resampling schemes have been proposed [27,28], such as multinomial resampling [1], residual resampling [29], stratified resampling [30], systematic resampling [31], and minimum variance resampling [32]. The branching particle method, another novel resampling scheme, was discussed in [33]. Reference [34] introduced four new branching PF that resolved the common complaint of unstable particle number and unpredictable results from the branching PF while reducing the computation complexity. The basic theory of the resampling technique involves discarding low-weighted particles and replicating high-weighted ones, thereby effectively alleviating the degeneracy problem.

All these improved techniques can be divided into two categories:

(1) those improving the proposal distribution to better approximate the real posterior-PDF. This technique enables more particles to be distributed where the real state is more likely to occur; and

(2) those using a resampling technique. By discarding low-weighted particles and replicating high-weighted ones, this technique can effectively alleviate the particle degeneracy problem.

Note that [18] point out a classical problem of the PF that always emerges in a special environment, specifically, when the likelihood function is much narrower than the prior. In this environment, most particles are discarded in the process of resampling as they have a low weight. This problem is exacerbated when the high probability region of the likelihood-PDF is far from that of the prior-PDF.

In our study, a new improved PF, termed the support vector regression PF (SVR-PF), was proposed to mainly overcome the problem in this special environment. The main procedure of this method is to fit the expression of the priori filtering density using SVR before obtaining the current observation. The particles then migrate to ensure they are distributed uniformly in a reasonable importance region (***IR***); the definition of ***IR*** is given in section III. Next, the priori and likelihood weights of each even-distributed particle is calculated according to the expression and observation equation, respectively. Then, the state variable is estimated by these even-distributed particles and their mix-weights, which are obtained by mixed and normalized a priori and likelihood weights, respectively. The main contributions arising from our study are:

(1) Particle degeneracy and impoverishment problems were re-discussed in detail using the concept of ***IR*** and particle density (*PD*), and subsequently used to evaluate the performance of the state estimation.

(2) Region amplification and the particle migration technique based on the SVR probability density estimation is proposed. In this procedure, particles are placed uniformly in the ***IR*** without changing the PDF of the original particles.

(3) Consistency between the proposed algorithm and Bayesian filtering is proved in theory.

(4) We verify the performance of our algorithm based on numerical simulations, the results from which suggest the effectiveness of the SVRPF, especially regarding narrow observation noise.

The outline of this paper is as follows. Section II introduces the basic principle of Bayesian filtering, the GPF, and other classic improved PFs. The drawback of GPF is re-discussed in detail and the motivation behind our idea is introduced in Section III. In Section IV, the SVR method used for estimating the probability density is introduced. In addition, the proposed algorithm is introduced in detail and its consistency with Bayesian filtering is proved from theory. An experiment is described in Section V that demonstrates the validity of the proposed algorithm. Section VI concludes this work.

## II. PRELIMINARIES

For the filtering algorithms introduced in this paper, the state space model is defined as [28]

$$\boldsymbol{x}_t = \mathbf{f}\left(\boldsymbol{x}_{t-1}, \boldsymbol{u}_{t-1}\right), \quad (1)$$

$$\boldsymbol{y}_t = \mathbf{h}\left(\boldsymbol{x}_t, \boldsymbol{v}_t\right), \quad (2)$$

where $\boldsymbol{x}_t \in \Re^{n_x}$ denotes the state variable at time step $t$, $n_x$ the dimension of the state vectors, and $\boldsymbol{y}_t \in \Re^{n_y}$ the observations; $\boldsymbol{u}_t \in \Re^{n_u}$ and $\boldsymbol{v}_t \in \Re^{n_v}$ denote the process noise and observation noise, respectively. We suppose that $\boldsymbol{u}_t$ and $\boldsymbol{v}_t$ are independent of each other, and the probability distribution of $\boldsymbol{u}_t$ and $\boldsymbol{v}_t$ are known. The mappings $\mathbf{f}: \Re^{n_x} \times \Re^{n_u} \mapsto \Re^{n_x}$ and $\mathbf{h}: \left(\Re^{n_x} \times \Re^{n_u}\right) \times \Re^{n_v} \mapsto \Re^{n_y}$ describe the system model and observation model, respectively; both are known. The system model is represented by a probabilistic form, $p\left(\boldsymbol{x}_t | \boldsymbol{x}_{t-1}\right)$, whereas the observation model is represented by $p\left(\boldsymbol{y}_t | \boldsymbol{x}_t\right)$. We also suppose the initial state $P\left(\boldsymbol{x}_0\right)$ is known.

### A. BAYESIAN FILTERING

The main purpose of a Bayesian estimation is to construct the posterior-PDF of a state variable using prior knowledge and observation. From Bayes' theorem, the posterior-PDF of $\boldsymbol{x}_{0:t}$ is expressed as

$$p\left(\boldsymbol{x}_{0:t} | \boldsymbol{y}_{1:t}\right) = \frac{p\left(\boldsymbol{y}_{1:t} | \boldsymbol{x}_{0:t}\right) p\left(\boldsymbol{x}_{0:t}\right)}{\int p\left(\boldsymbol{y}_{1:t} | \boldsymbol{x}_{0:t}\right) p\left(\boldsymbol{x}_{0:t}\right) d\boldsymbol{x}_{0:t}}, \quad (3)$$

where $p\left(\boldsymbol{x}_{0:t}\right)$ denotes the prior probability distribution, $p\left(\boldsymbol{y}_{1:t} | \boldsymbol{x}_{0:t}\right)$ the likelihood-PDF, and $p\left(\boldsymbol{x}_{0:t} | \boldsymbol{y}_{1:t}\right)$ the posterior-PDF.

The prior-PDF $p\left(\boldsymbol{x}_t | \boldsymbol{y}_{1:t-1}\right)$ was used to replace $p\left(\boldsymbol{x}_{0:t}\right)$ so as to accomplish this state variable estimation recursively. The posterior-PDF is estimated using a recursive method:

1) *Prediction*—calculate the prior-PDF at the current time according to the state model and the posterior-PDF $p\left(\boldsymbol{x}_{t-1} | \boldsymbol{y}_{1:t-1}\right)$ at time step $t-1$.

2) *Update*—calculate the posterior-PDF at the current time according to the observation $\boldsymbol{y}_t$ and the prior-PDF.

In detail, the prior-PDF and the posterior-PDF at the current time is given by

$$p\left(\boldsymbol{x}_t \mid \boldsymbol{y}_{1:t-1}\right)=\int p\left(\boldsymbol{x}_t \mid \boldsymbol{x}_{t-1}\right) p\left(\boldsymbol{x}_{t-1} \mid \boldsymbol{y}_{1:t-1}\right) d\boldsymbol{x}_t , \tag{4}$$

$$p\left(\boldsymbol{x}_t \mid \boldsymbol{y}_{1:t}\right)=\frac{p\left(\boldsymbol{y}_t \mid \boldsymbol{x}_t\right) p\left(\boldsymbol{x}_t \mid \boldsymbol{y}_{1:t-1}\right)}{p\left(\boldsymbol{y}_t \mid \boldsymbol{y}_{1:t-1}\right)}, \tag{5}$$

where

$$p\left(\boldsymbol{y}_t \mid \boldsymbol{y}_{1:t-1}\right)=\int p\left(\boldsymbol{y}_t \mid \boldsymbol{x}_t\right) p\left(\boldsymbol{x}_t \mid \boldsymbol{y}_{1:t-1}\right) d\boldsymbol{x}_t . \tag{6}$$

The state estimation for the current time is obtained by implementing the *prediction* and *update* steps repeatedly. However, there are but few systems that can be solved by this analytic solution directly as the evaluation of the integral in (4) is difficult.

### B. GENERAL PF

To avoid the difficult integration for the nonlinear and non-Gaussian systems, i.e., the general PF (GPF) or so-called bootstrap filter algorithm [4], an approximate Bayesian filtering method was developed based on simulations exploiting the Monte Carlo principle [35].

Replacing the calculation of the prior-PDF using integration directly, the GPF simulates the integration by propagating particles $\left(\boldsymbol{x}_t^i\right)_{i=1,\cdots,N}$ and their weights $\left(w_t^i\right)_{i=1,\cdots,N}$, where $N$ denotes the number of particles. In brief, the GPF procedure is as follows:

(1) *Initialization*—begin by sampling particles $\left(\boldsymbol{x}_0^i\right)_{i=1,\cdots,N}$ from $p\left(\boldsymbol{x}_0\right)$ and then set the weights of the initial particles to $\left(w_0^i\right)_{i=1,\cdots,N}=N^{-1}$; for convenience, particles and their weights are composited $\left\{\boldsymbol{x}_0^i, w_0^i\right\}$.

(2) *Prediction* (step $t$)—calculate $\hat{\boldsymbol{x}}_t^i$ according to the state model and $\left\{\boldsymbol{x}_{t-1}^i, w_{t-1}^i\right\}$; after this stage, particles and their weights can be expressed as $\left\{\hat{\boldsymbol{x}}_t^i, w_{t-1}^i\right\}$.

(3) *Update*—calculate likelihood weights $\tilde{w}_t^i$ according to the observation model and $\boldsymbol{y}_t$; particles and their weights become $\left\{\hat{\boldsymbol{x}}_t^i, \tilde{w}_t^i\right\}$.

(4) *Mixed weights and normalization*—

$$\hat{w}_t^i=\frac{\tilde{w}_t^i w_{t-1}^i}{\sum_{i=1}^{N} \tilde{w}_t^i w_{t-1}^i}; \tag{7}$$

after this stage, particles and their weights are $\left\{\hat{\boldsymbol{x}}_t^i, \hat{w}_t^i\right\}$.

(5) *State estimation*—

$$\overline{\boldsymbol{x}}_t=\sum_{i=1}^{N} \hat{w}_t^i \hat{\boldsymbol{x}}_t^i ; \tag{8}$$

(6) *Systematic Resampling*—

Set $\Omega_0$=0 and $\Omega_n=\sum_{i=1}^{n} \hat{w}_t^i \quad (n=1,\cdots,N)$

for $i$=1, …, $N$

Draw [0,1] -uniform $U_i$

While $\left(U_i \in\left(\Omega_{j-1}, \ \Omega_j\right] \ (j=1,\cdots N)\right)$

Set $\boldsymbol{x}_t^i=\hat{\boldsymbol{x}}_t^j \quad w_t^i=N^{-1}$

End for

(7) *Reiterate*—repeat steps 2–6 until the end.

The correspondence between the GPF and Bayesian filtering is understood as follows:

During the execution of the GPF, $\left\{\boldsymbol{x}_0^i, w_0^i\right\}$ in step 1 is used to describe the initial state distribution; $\left\{\hat{\boldsymbol{x}}_t^i, w_{t-1}^i\right\}$ in step 2 is used to describe the prior-PDF referred to in (4); $\left\{\hat{\boldsymbol{x}}_t^i, \tilde{w}_t^i\right\}$ in step 3 is used to describe the likelihood-PDF denoted by $p\left(\boldsymbol{y}_t \mid \boldsymbol{x}_t\right)$; $\left\{\hat{\boldsymbol{x}}_t^i, \hat{w}_t^i\right\}$ in step 4 is used to describe the posterior-PDF referred to in (5); the resampling step is used to mitigate the degeneracy problem referred to in (6). As with $\left\{\hat{\boldsymbol{x}}_t^i, \hat{w}_t^i\right\}$, $\left\{\boldsymbol{x}_t^i, w_t^i\right\}$ is also used to describe the posterior-PDF. Compared with $\left\{\hat{\boldsymbol{x}}_t^i, \hat{w}_t^i\right\}$, $\left\{\boldsymbol{x}_t^i, w_t^i\right\}$ was obtained using the systematic resampling technique. This resampling stage removes those particles with low weights and repeats the particles with high weights, thereby avoiding numerous invalid calculations while not changing the discrete distribution supported by $\left\{\hat{\boldsymbol{x}}_t^i, \hat{w}_t^i\right\}$. This systematic resampling technique was demonstrated to be unbiased in [36].

### C. OTHER CLASSIC IMPROVED PFS

At present, the improvement of PF is mainly based on the improvement of the proposal distribution and the resampling technique. However, improving the proposal distribution often needs the cooperation of the resampling technique. In this section, several classic improved PFs are introduced.

1 The extended Kalman PF (EPF) and the unscented Kalman PF (UPF)—The core idea of the EPF is to modify the proposal distribution using the EKF. The procedure for the EPF is in summary [20]:

(1) Initialization—begin by sampling particles $\left(\boldsymbol{x}_0^i\right)_{i=1,\cdots,N}$ from $p\left(\boldsymbol{x}_0\right)$ and then set the weights of the initial particles to $\left(w_0^i\right)_{i=1,\cdots,N}=N^{-1}$. Set the initial state noise covariance matrix $\boldsymbol{P}_0^i$ and the initial observation covariance matrix $\boldsymbol{Q}_0^i$

(2) Predict the update state of each particle using EKF:

$$\overline{\boldsymbol{x}}_{t,pre}^i=\mathbf{f}\left(\boldsymbol{x}_{t-1}^i\right) \tag{9}$$

$$\boldsymbol{P}_{t,pre}^i=\boldsymbol{F}_t^i P_{t-1,pre}^i\left(\boldsymbol{F}_t^i\right)^T+\boldsymbol{Q}_t^i \tag{10}$$

$$\boldsymbol{K}_t^i=P_{t,pre}^i\left(\boldsymbol{H}_t^i\right)^T\left[\boldsymbol{R}_t+\boldsymbol{H}_t^i P_{t,pre}^i\left(\boldsymbol{H}_t^i\right)^T\right]^{-1} \tag{11}$$

$$\overline{\boldsymbol{x}}_t^i=\overline{\boldsymbol{x}}_{t,pre}^i+\boldsymbol{K}_t^i\left[\boldsymbol{y}_t-\boldsymbol{h}\left(\overline{\boldsymbol{x}}_{t,pre}^i\right)\right], \tag{12}$$

$$\boldsymbol{P}_t^i = \boldsymbol{P}_{t,pre}^i - \boldsymbol{K}_t^i \boldsymbol{H}_t^i \boldsymbol{P}_{t,pre}^i, \tag{13}$$

where $\boldsymbol{F}_t^i = \nabla_x \mathbf{f}(\boldsymbol{x}, \boldsymbol{u}_{t-1})\big|_{\boldsymbol{x}=\boldsymbol{x}_{t-1}^i}$ and $\boldsymbol{H}_t^i = \nabla_x \mathbf{h}(\boldsymbol{x}_t, \boldsymbol{v}_t)\big|_{\boldsymbol{x}=\bar{\boldsymbol{x}}_{t,pre}^i}$.

(3) Update particles, then calculate and normalize the mixed weights

$$\hat{\boldsymbol{x}}_t^i \sim q\left(\boldsymbol{x}_t^i \middle| \boldsymbol{x}_{0:t-1}^i, \boldsymbol{y}_{1:k}\right) = N\left(\bar{\boldsymbol{x}}_t^i, \boldsymbol{P}_t^i\right), \tag{14}$$

$$\tilde{w}_t^i \propto p\left(\boldsymbol{y}_k \middle| \hat{\boldsymbol{x}}_t^i\right) p\left(\hat{\boldsymbol{x}}_t^i \middle| \boldsymbol{x}_{t-1}^i\right), \tag{15}$$

$$\hat{w}_t^i = \frac{\tilde{w}_t^i}{\sum_{i=1}^N \tilde{w}_t^i}; \tag{16}$$

(4) State estimation

$$\bar{\boldsymbol{x}}_t = \sum_{i=1}^N \hat{w}_t^i \hat{\boldsymbol{x}}_t^i; \tag{17}$$

(5) *Systematic Resampling*

Set $\Omega_0$=0 and $\Omega_n = \sum_{i=1}^n \hat{w}_t^i \quad (n = 1, \cdots, N)$

for $i$=1,…,$N$

Draw [0,1] -uniform $U_i$

While ($U_i \in \left(\Omega_{j-1}, \ \Omega_j\right] \ (j = 1, \cdots N)$) set $\boldsymbol{x}_t^i = \hat{\boldsymbol{x}}_t^j$

Set $w_t^i = N^{-1}$

End for

(6) *Reiterate*—repeat steps 2–6 until the end.

The difference between UPF and EPF amounts to replacing EKF with UKF (see [37])

2 Minimum variance resampling

Minimum variance resampling was introduced in [32] and was used to replace step (6) in the GPF procedure. Details of the minimum variance resampling are summarized in [34]:

Initialization: $\mathbb{N}_t^N := N, W := N$

Repeat: for $i$ =1,2, …, $N$-1 do

$\mathbb{N}_t^i := \left\lfloor N\hat{w}_t^i \right\rfloor$

Draw [0,1] – uniform $U_i$

If $\left\{N\hat{w}_t^i\right\} + \left\{W - N\hat{w}_t^i\right\} < 1 \ \ \& \ \ U_i \le \frac{\left\{N\hat{w}_t^i\right\}}{\{W\}}$ then,

$$\mathbb{N}_t^i := \mathbb{N}_t^i + \mathbb{N}_t^N - \lfloor W \rfloor \tag{18}$$

else if $\left\{N\hat{w}_t^i\right\} + \left\{W - N\hat{w}_t^i\right\} \ge 1 \ \ \& \ \ U_i \ge \frac{\left\{N\hat{w}_t^i\right\} - \{W\}}{1 - \{W\}}$

then

$$\mathbb{N}_t^i := \mathbb{N}_t^i + \mathbb{N}_t^N - \lfloor W \rfloor \tag{19}$$

else if $\left\{N\hat{w}_t^i\right\} + \left\{W - N\hat{w}_t^i\right\} \ge 1 \ \ \& \ \ U_i < \frac{\left\{N\hat{w}_t^i\right\} - \{W\}}{1 - \{W\}}$,

then,

$$\mathbb{N}_t^i := \mathbb{N}_t^i + 1 \tag{20}$$

For $k = 0, 1, \cdots, \mathbb{N}_t^i - 1$ do $\boldsymbol{x}_t^{\mathbb{N}_t^i - k} := \hat{\boldsymbol{x}}_t^i$

$$\mathbb{N}_t^N := \mathbb{N}_t^N + \mathbb{N}_t^i, W := W - N\hat{w}_t^i \tag{21}$$

Set $w_t^i = N^{-1}$

End for

3 Residual resampling

Applying the residual resampling technique separates the decimal part of $\left\{N\hat{w}_t^i\right\}$ for resampling and ensures fewer particles participate in the redistribution process and generates less resampling noise. The outline of the residual resampling procedure is:

Initialization: $N_{res} := N - \sum_{i=1}^N \left\lfloor N\hat{w}_t^i \right\rfloor, \ \ \tilde{U}_0 = m = i = 1$

For $k = 1, 2, \cdots, N_{res}$

Draw [0,1] – uniform $U_k$ and Set $\tilde{U}_k = U_k^{\frac{1}{N-k+1}} \tilde{U}_{k-1}$

While ($\tilde{U}_k > \sum_{i=1}^m \frac{N\hat{w}_t^i - \left\lfloor N\hat{w}_t^i \right\rfloor}{N_{res}}$), Set $m$=$m$+1

$\boldsymbol{x}_t^k = \hat{\boldsymbol{x}}_t^m$

End for

While ($i \le N$)

If $\left\lfloor N\hat{w}_t^i \right\rfloor \ge 1$

Set $\boldsymbol{x}_t^{N_{res}+1:N_{res}+\left\lfloor N\hat{w}_t^i \right\rfloor} = \hat{\boldsymbol{x}}_t^i$, $N_{res} = N_{res} + \left\lfloor N\hat{w}_t^i \right\rfloor$,

$i = i + 1$.

Set $w_t^i = N^{-1} \ \left(i = 1, 2, \cdots N\right)$

4 Branching PF

The resampling technique of the branching PF is different from the above two resampling methods. Its core idea is to preserve some part of the particles and their weights, and redistribute other particles and their weights. In the process of redistribution, the number of particles cannot be fixed. The procedure of the resampling step for the branching PF is in outline as follows:

Initialization $\mathbb{N}_t = 0$.

For $i$=1,2,…, $N$

If $N\hat{w}_t^i \notin (a_t, b_t)$, where $0 < a_t < b_t < N$

$\mathbb{N}_t^i := \left\lfloor N\hat{w}_t^i \right\rfloor + \rho_t^i$, with $\rho_t^i$ a $\left(N\hat{w}_t^i - \left\lfloor N\hat{w}_t^i \right\rfloor\right)$ -Bernoulli

Set $w_t^{\mathbb{N}_t+1:\mathbb{N}_t+\mathbb{N}_t^i} = N^{-1}$, $\boldsymbol{x}_t^{\mathbb{N}_t+1:\mathbb{N}_t+\mathbb{N}_t^i} = \hat{\boldsymbol{x}}_t^i$

$$\mathbb{N}_t = \mathbb{N}_t + \mathbb{N}_t^i \tag{22}$$

else

$$w_t^i = \hat{w}_t^i, \ \boldsymbol{x}_t^i = \hat{\boldsymbol{x}}_t^i, \ \mathbb{N}_t = \mathbb{N}_t + 1 \tag{23}$$

End for.

The improvement of the proposal distribution (such as EPF and UPF) enables more particles to be concentrated where the real state is more likely to occur. However, approaching the real posterior-PDF accurately is difficult in nonlinear and non-Gaussian environments. In these circumstances, the degree of approximation between the proposal distribution and the posterior-PDF is directly related to the accuracy of the state estimation, and this accuracy is more sensitive to the selection of the proposal distribution in a special environment: when the likelihood function is much narrower than the prior. Moreover, improvements of the proposal distribution usually require the introduction of a current observation, which might increase the calculated amount thereafter and influence the time-effectiveness of the algorithm.

The resampling technique does not directly improve the state estimation accuracy at the current time. Indeed, [34] advocates executing the state estimation stage before resampling stage in the PF algorithm to avoid introducing the error caused by resampling. Furthermore, in the framework of general PF, any resampling method simply repeats or deletes particles based on the prior-PDF. In the special environment, the resampling technique does not guarantee that more particles are obtained in the high probability region of the likelihood-PDF, which might greatly reduce the accuracy of the state estimation especially when this region of the likelihood-PDF is far from that of the prior-PDF.

## III. MOTIVATION

The advantage of the PF is that by propagating numerous particles the nonlinear transformation of an arbitrary distribution for any nonlinear system is determined accurately. However, with only a finite number of particles, describing perfectly a continuous probability distribution is impossible. This basic fact underscores the degeneracy and impoverishment problems.

This section mainly discusses the effects of particle degeneracy and the impoverishment problem on the performance of the PF in the special environment, when the likelihood function is much narrower than the prior. The motivation of the proposed algorithm is then introduced.

### A. PROBLEM STATEMENT

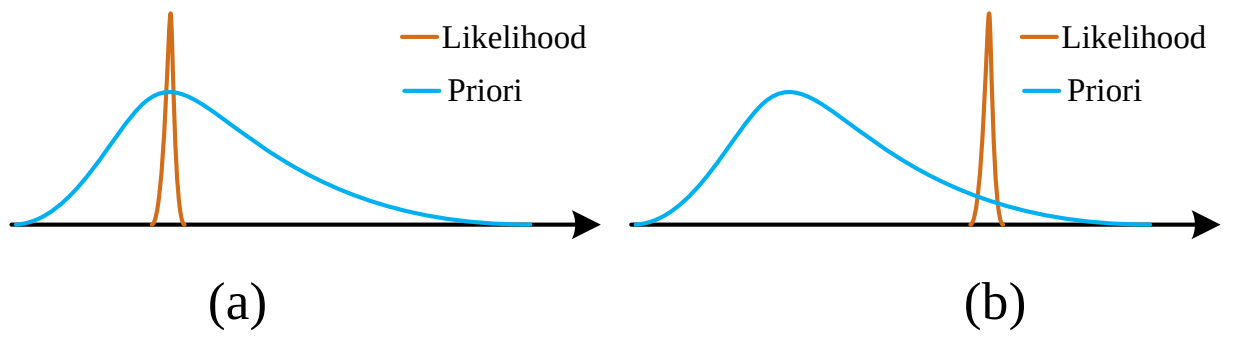

FIGURE 1. Narrow observation noise: (a) likelihood function coincides with prior; (b) likelihood function far from prior

When the likelihood function is much narrower than the prior, two situations arise (Fig. 1): (a) the region of high likelihood coincides with the region of high prior and (b) the region of high likelihood is located in one of the tails of the prior density.

For situation (a), the particles generated from the proposal distribution are located in the region of high likelihood. The proposal distribution is then represented by the distribution of particles, and the likelihood function can be described by these particles and their likelihood weights. Finally, state estimation is calculated using these particles and their mixed weights. However, the weights of the low-weight particles will be lower and the weights of the high-weight particles will gradually spread to the low-weight particles in the recursion-estimation process. This phenomenon forces more particles to lose efficacy in state estimation and leads to particle degeneracy. Fortunately, by deleting low-weight particles and replicating high-weight particles, the resampling technique ensures that particles are concentrated in the region of high posterior and overcomes the particle degeneracy problem well, although the resampling technique may introduce some errors.

For situation (b), when the priori distribution is used as the proposal distribution (GPF), most particles are far from the region of high likelihood. At this point, these particles cannot represent the likelihood distribution integrally and accurately, which might result in the loss of likelihood information and this problem can be seen clearly in section V Fig. 6. To solve this problem, [18] proposed the EPF, which takes the estimated results of EKF as the proposal distribution. The core idea of EPF is to concentrate particles in the region of high approximated posterior-PDF, with the approximated posterior-PDF estimated by the EKF. This method effectively improves the performance of state estimation in the narrow observation noise environment. However, the narrow likelihood function is far away from the prior distribution, which indicates $\boldsymbol{y}_t - \boldsymbol{h}\left(\bar{\boldsymbol{x}}_{t,pre}^{i}\right)$ (the last term in (12)) is larger than that in situation (a). Actually, the truncation error of EKF derives from this last term implying that the magnitude of the truncation error is positively correlated with the magnitude of $\boldsymbol{y}_t - \boldsymbol{h}\left(\bar{\boldsymbol{x}}_{t,pre}^{i}\right)$. Moreover, the narrow likelihood generates a narrow posterior-PDF, which result in particles estimated using (12) to concentrate in a narrow region. A larger truncation error is likely to cause the particles to deviate from the region of high likelihood as the likelihood is narrow. In other words, in situation (b), EKF concentrates particles in the narrow region but deviates from the region of high likelihood, resulting in these impoverishment problems. This phenomenon has been verified in Fig. 6. Figs. 6 and 7 show that the EPF does effectively improve the performance of state estimation in the narrow observation noise environment, but truncation errors are also introduced. The resampling stage usually occurs after the calculation of mixed weights. Therefore, the resampling technique cannot solve the loss of likelihood information for the reason that the likelihood function is far from the prior.

### B. DISCUSSION OF THE NEW PERSPECTIVE FOR THE PARTICLE FILTER

To facilitate the discussion of this special problem caused by the narrow observation noise, the concepts of importance region (***IR***) and particle density (*PD*) are defined as follows:

***Definition*** **1**: Suppose $\boldsymbol{X}$ is a $d \times N$ matrix, each of its columns being a sample from the multidimensional distribution function $F(\boldsymbol{x})$, where the dimension of $\boldsymbol{x}$ is $d$. Let $\boldsymbol{x}_{real}$ be the real state at the current time that is generated from $F(\boldsymbol{x})$; let $\boldsymbol{x}_{\max}$ and $\boldsymbol{x}_{\min}$ be two $d \times 1$ vectors, each of their rows being equal to the maximum and minimum, respectively, of each row of $\boldsymbol{X}$. Then ***IR*** is defined as the $d \times 2$ matrix for which

$$\boldsymbol{IR} = \left[\boldsymbol{x}_{\min}, \boldsymbol{x}_{\max}\right], \tag{23}$$

called the *importance region* (***IR***), each row of ***IR*** represents the important region of $\boldsymbol{X}$ in that dimension, and

$$PD = \frac{\prod_{i=1}^{d}\left[\boldsymbol{x}_{\max}(i) - \boldsymbol{x}_{\min}(i)\right]}{N} \tag{24}$$

is called the *particle density* (*PD*).

Suppose that the proposal distribution is $F(\boldsymbol{x})$ in Definition 1, and $\boldsymbol{x} \in \Re^{n_x}$. Then ***IR*** describing the region of particles and *PD* describing the density of particles in the ***IR*** may be used to evaluate the performance of the PF. In addition, ***IR*** and *PD* increase as $N$ increases and when $N \to \infty$, then $IR \to \Re^{n_x}$ and $PD \to \infty$. However, the rate of increase in ***IR*** and *PD* decreases sharply with increasing *N*, which prevents us from blindly increasing the particle number to improve the estimation performance.

When the particle number is fixed, increasing ***IR*** allows the particles to describe the proposal distribution more integrally. However, increasing ***IR*** also decreases *PD*, so that the particles cannot represent the proposal distribution accurately. Hence, the conflict between integrity and accuracy is embodied in ***IR*** and *PD*. Nonetheless, particle degeneracy and the impoverishment problem can be explained by ***IR*** and *PD* as follows:

**Remark 1.** *The particle degeneracy problem is due to an ampliative **IR** during the iteration thereby reducing the PD and affects the accuracy of fitting of the posteriori-PDF.*

**Remark 2.** *The impoverishment problem arises because the **IR** is too narrow, hence the integrity of the posterior-PDF is lost.*

During the whole PF procedure, particles should not only be able to describe the proposal distribution, but also be able to describe the likelihood distribution with their likelihood weights. That requires a sufficient number of particles to be located in the region of high likelihood. However, in situation (b), it is difficult to sample enough particles in the region of high likelihood without considering the current observation. EPF improves the performance of state estimation in this special environment, but introduces truncation errors. In addition, the introduction of the current observation to improve the proposal distribution may increase the calculated amount after obtaining the current observation and influence the time-effectiveness of the algorithm.

The problem to be solved in this work is, in circumstances when the current observation is not obtained, and the likelihood is far from the prior, how can we ensure that there are still enough particles in the region of high likelihood? To the best of our knowledge, the ***IR*** of existing PF algorithms is determined based on particles generated by the proposal distribution, and particles are always concentrated in the region of high proposal distribution. However, the region of high likelihood is not always in the region of high proposal distribution. Therefore, setting particles in an appropriate ***IR*** uniformly has become the motivation of our work (Fig. 2). Meanwhile, the SVR technique is used to update the prior weight, which ensures a uniform dispersal of particles [Fig. 2(b)] and their weights represent the same distribution as particles in Fig. 2(a).

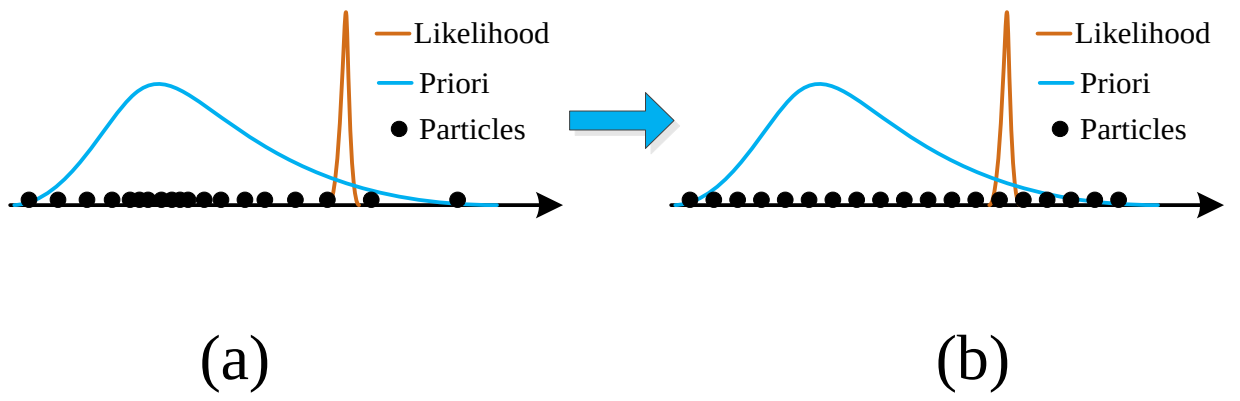

**FIGURE 2. Motivation of our method (a) the distribution of particles in traditional PF; (b) the distribution of particles in our method**

## IV. PROPOSED ALGORITHM

In this section, SVR is introduced as a probability density fitting tool. Then, region amplification and the particle migration technique are presented in detail along with the SVRPF algorithm. Finally, the consistency between the SVRPF and Bayesian filtering is demonstrated step by step in theory.

### *A. SVR PROBABILITY DENSITY ESTIMATION*

As a regression algorithm [38,39], SVR is widely used as its performance in small sample nonparametric estimations is excellent. Here, it can be used to estimate the probability density [40,41].

With the priori-PDF denoted by $\left\{\hat{\boldsymbol{x}}_t^i, w_{t-1}^i\right\}$, its expression may be estimated using the SVR. The empirical distribution function has expansion

$$F(\boldsymbol{x}_t) = \sum_{i=1}^{N} w_{t-1}^i \theta\left(\boldsymbol{x}_t \succ \hat{\boldsymbol{x}}_t^i\right), \tag{25}$$

where $\theta\left(\boldsymbol{x}_t \succ \hat{\boldsymbol{x}}_t^i\right)$ satisfies

$$\theta\left(\boldsymbol{x}_t \succ \hat{\boldsymbol{x}}_t^i\right) = \begin{cases} 1, & \boldsymbol{x}_t(j) > \hat{\boldsymbol{x}}_t^i(j) \quad j = 1, 2 \cdots D \\ 0, & otherwise \end{cases}, \tag{26}$$

with $D$ denoting the dimension of $\hat{\boldsymbol{x}}_t^i$. Then, SVR is used to process the data

$$\left\{\left[\hat{\boldsymbol{x}}_t^1, F\left(\hat{\boldsymbol{x}}_t^1\right)\right], \left[\hat{\boldsymbol{x}}_t^2, F\left(\hat{\boldsymbol{x}}_t^2\right)\right], \cdots \left[\hat{\boldsymbol{x}}_t^N, F\left(\hat{\boldsymbol{x}}_t^N\right)\right]\right\} \tag{27}$$

to construct an estimation of the PDF,

$$p\left(\boldsymbol{x}_t,\boldsymbol{\beta}_t\right)=\sum_{i=1}^{N}\beta_{t,i}k\left(\boldsymbol{x}_t,\hat{\boldsymbol{x}}_t^i\right), \tag{28}$$

where $\boldsymbol{\beta}_t$ is a set of parameters of the estimation function, and $k(\bullet)$ the kernel function. The estimation of the cumulative distribution function (CDF) is obtained from

$$F\left(\boldsymbol{x}_t\right)=\int_{-\infty}^{x_t}p\left(\boldsymbol{u},\boldsymbol{\beta}_t\right)d\boldsymbol{u}, \tag{29}$$

$$=\sum_{i=1}^{N}\beta_{t,i}K\left(\boldsymbol{x}_t,\hat{\boldsymbol{x}}_t^i\right), \tag{30}$$

where $K\left(\boldsymbol{x}_t,\hat{\boldsymbol{x}}_t^i\right)$ satisfies

$$K\left(\boldsymbol{x}_t,\hat{\boldsymbol{x}}_t^i\right)=\int_{-\infty}^{x_t}k\left(\boldsymbol{u},\hat{\boldsymbol{x}}_t^i\right)d\boldsymbol{u}. \tag{31}$$

Being different from the general SVR, and to satisfy the property of a PDF, $k\left(\boldsymbol{x}_t,\hat{\boldsymbol{x}}_t^i\right)$ and $\boldsymbol{\beta}_t$ must be restricted,

$$k\left(\boldsymbol{x}_t,\hat{\boldsymbol{x}}_t^i\right)\geq 0 \;\; and \;\; \int_{-\infty}^{+\infty}k\left(\boldsymbol{x}_t,\hat{\boldsymbol{x}}_t^i\right)d\boldsymbol{x}_t=1, \tag{32}$$

$$\sum_{i=1}^{N}\beta_{t,i}=1. \tag{33}$$

In the SVR algorithm, there are several kinds of kernel functions. Different kernel functions ensure the PDF obtained by the SVR algorithm have slight differences, and the selection of kernel function is usually based on the engineering or experimental test [42–45]. In this work, we focus on developing a SVR algorithm into a GPF to form a new PF framework to overcome the problem caused by narrow observation noise. Therefore, the selection of kernel functions is not discussed in this paper, but $k(\bullet)$ and $K(\bullet)$ are assumed to take the mathematical forms

$$k\left(\boldsymbol{x}_t,\hat{\boldsymbol{x}}_t^i\right)=\prod_{j=1}^{D}\frac{\lambda}{2+e^{\lambda\left(x_t(j)-\hat{x}_t^i(j)\right)}+e^{-\lambda\left(x_t(j)-\hat{x}_t^i(j)\right)}}, \tag{34}$$

$$K\left(\boldsymbol{x}_t,\hat{\boldsymbol{x}}_t^i\right)=\prod_{j=1}^{D}\frac{1}{1+e^{-\lambda\left(x_t(j)-\hat{x}_t^i(j)\right)}}, \tag{35}$$

respectively. The $\varepsilon-SVR$ [46] is used to solve this quadratic programming problem, the object function and restraint satisfying expressions

$$\min\;\sum_{i=1}^{N}\sum_{i'=1}^{N}\beta_{t,i}\beta_{t,i'}k\left(\hat{\boldsymbol{x}}_t^i,\hat{\boldsymbol{x}}_t^{i'}\right)+C\sum_{i=1}^{N}\left(\xi_t^{i'}+\xi^{*i'}_t\right), \tag{36}$$

$$s.t.\;\; F\left(\hat{\boldsymbol{x}}_t^{i'}\right)-\sum_{i=1}^{N}\beta_{t,i}K\left(\hat{\boldsymbol{x}}_t^{i'},\hat{\boldsymbol{x}}_t^i\right)\leq\varepsilon+\xi^{*i'}_t \quad i'=1,2,\cdots N, \tag{37}$$

$$\sum_{i=1}^{N}\beta_{t,i}K\left(\hat{\boldsymbol{x}}_t^{i'},\hat{\boldsymbol{x}}_t^i\right)-F\left(\hat{\boldsymbol{x}}_t^{i'}\right)\leq\varepsilon+\xi_t^{i'} \quad i'=1,2,\cdots N, \tag{38}$$

$$\sum_{i=1}^{N}\beta_{t,i}=1, \tag{39}$$

$$\xi_t^{i'}\geq 0,\xi^{*i'}_t\geq 0,\beta_{t,i}\geq 0 \quad i'=1,2,\cdots N, \tag{40}$$

where $C$ denotes the penalty factor; $\xi_t^{i'}$ and $\xi^{*i'}_t$ denote slack variables, and

$$\varepsilon=\min\sqrt{\frac{1}{N}F\left(\hat{\boldsymbol{x}}_t^{i'}\right)\left(1-F\left(\hat{\boldsymbol{x}}_t^{i'}\right)\right)} \quad i'=1,2,\cdots N. \tag{41}$$

The parameters $\boldsymbol{\beta}_t$ of this quadratic programming problem can be solved. Subsequently, the estimations of PDF and CDF are obtained from (14) and (16), respectively.

### B. REGION AMPLIFICATION AND PARTICLES MIGRATION

The expression of the proposal distribution is obtained using the SVR and is used in turn to calculate the SVR weights of these uniformly-dispersed particles. The core idea of this innovation is that these new particles and their SVR weights describe also the proposal distribution. The advantage of this procedure is that particle positions are made more flexible and an excessive dependence on the proposal distribution is avoided.

*The **IR** selection principle*—The ***IR*** should contain most regions of the posterior-PDF to determine the posterior-PDF in a more integrated manner. For finite numbers of particles, the IR should not be too large so as to avoid a PD that is too low and should reflect the posteriori-PDF accurately.

In essence, the ***IR*** selection principle balances the integrity and accuracy of the posterior-PDF with a finite number of particles. The size of the ***IR*** and the PD reflect the integrity and accuracy of the posterior-PDF, respectively. In the GPF, the proposal distribution is the prior-PDF. The GPF-***IR*** contains roughly but not accurately the posterior-PDF. We compensate this poor quality by amplifying the GPF-***IR***. We call the amplified region the SVR-***IR***.

Suppose $\left[\boldsymbol{x}_{GPF,\min},\boldsymbol{x}_{GPF,\max}\right]$ is a GPF-***IR***; the SVR-***IR*** is selected as follows,

$$\left[\boldsymbol{x}_{SVR,\min},\;\boldsymbol{x}_{SVR,\max}\right]=\left[\boldsymbol{x}_{GPF,\min}-\eta\boldsymbol{\Delta},\;\boldsymbol{x}_{GPF,\max}+\eta\boldsymbol{\Delta}\right], \tag{42}$$

where $\boldsymbol{\Delta}=\boldsymbol{x}_{GPF,\max}-\boldsymbol{x}_{GPF,\min}$, $\eta$ is an amplification factor, and $\eta>0$. The role of $\eta$ is to balance the integrity and accuracy of the posterior-PDF. After that, the particle migration technique is used to disperse particles uniformly in the SVR-***IR***.

In the GPF, $\left\{\hat{\boldsymbol{x}}_t^i,w_{t-1}^i\right\}$ represents the prior-PDF

$$p\left(\boldsymbol{x}_t\left|\boldsymbol{y}_{1:t-1}\right.\right)\leftarrow\left\{\hat{\boldsymbol{x}}_t^i,w_{t-1}^i\right\}. \tag{43}$$

Now, $\left\{\hat{\boldsymbol{x}}_t^i,w_{t-1}^i\right\}$ is used in fitting the prior-PDF

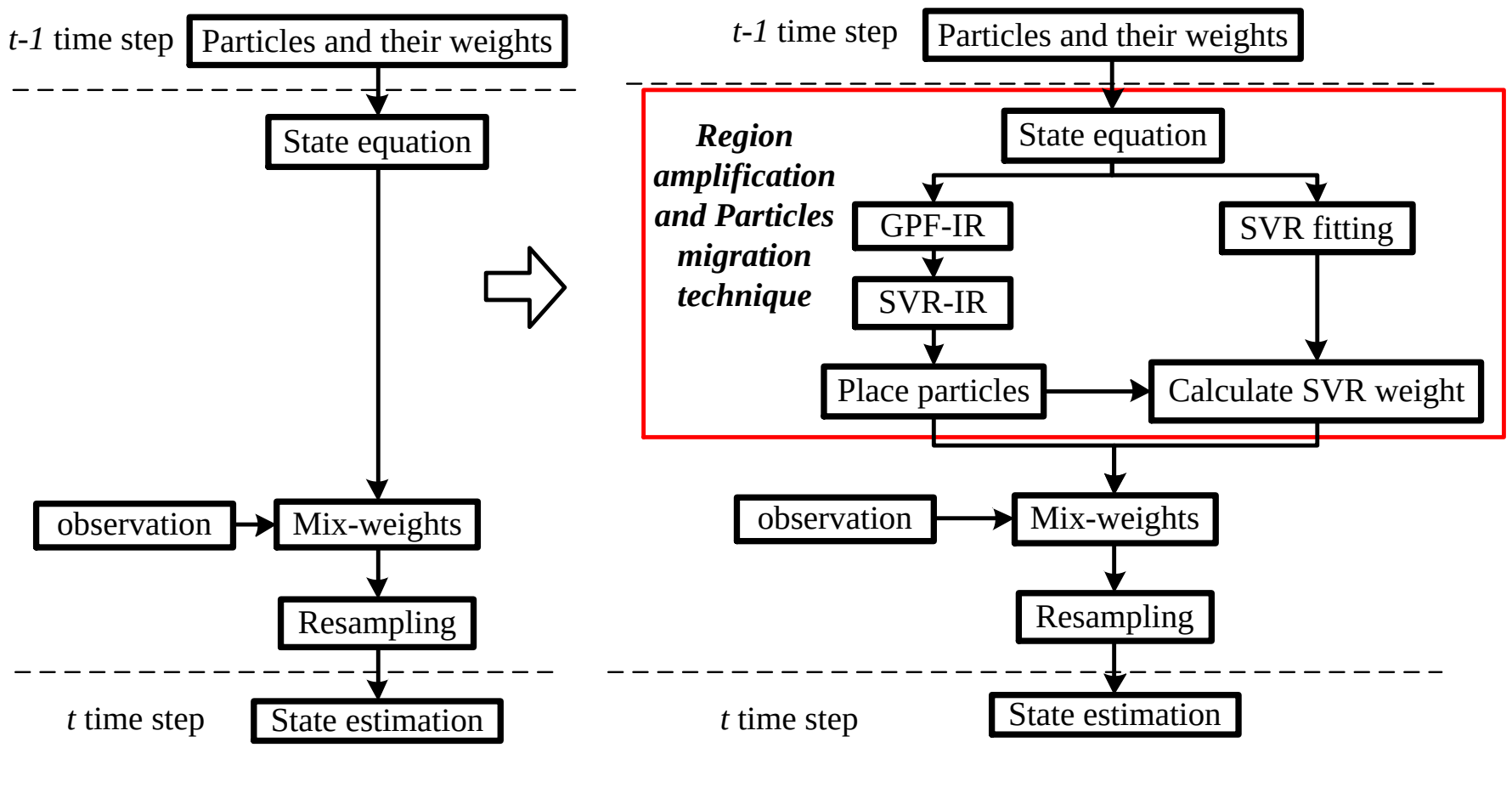

**FIGURE. 3 Integral structures of the GPF and SVRPF.**

**TABLE I**
REGION AMPLIFICATION AND PARTICLE MIGRATION

| |
|---|
| Input: $\left\{\hat{\boldsymbol{x}}_t^i, w_{t-1}^i\right\}_{i=1,\cdots,N}$ |
| Output: $\left\{\hat{\boldsymbol{x}}_{SVR,t}^j, \bar{w}_{SVR,t-1}^j\right\}_{j=1,\cdots,M}$, $M \ge N$ |
| (1) Calculate $\boldsymbol{\beta}_t$ by solving the quadratic programming problem of (36)–(40) |
| (2) Select the SVR-***IR*** according to (42) |
| (3) Place $M$ particles uniformly in SVR-***IR***. |
| (4) Calculate the SVR weights according to (45) and (49) |
| (5) Get new particles and their weights according to (47) |

$p\left(\boldsymbol{x}_t \middle| \boldsymbol{y}_{1:t-1}\right)$ using the SVR, the expression being

$$p\left(\boldsymbol{x}_t \middle| \boldsymbol{y}_{1:t-1}\right) \approx \sum_{i=1}^{N} \beta_{t,i} K\left(\boldsymbol{x}_t, \hat{\boldsymbol{x}}_t^i\right). \tag{44}$$

We next place $M$ particles uniformly in SVR-***IR*** and calculate the SVR weights,

$$w_{SVR,t-1}^j = \sum_{i=1}^{N} \beta_{t,i} k\left(\hat{\boldsymbol{x}}_{SVR,t}^j, \hat{\boldsymbol{x}}_t^i\right) \quad j = 1, 2 \cdots, M . \tag{45}$$

Normalizing the SVR weights, we have

$$\bar{w}_{SVR,t-1}^j = w_{SVR,t-1}^j / \sum_{j=1}^{M} w_{SVR,t-1}^j . \tag{46}$$

The new particles and their weights are

$$\left\{\hat{\boldsymbol{x}}_{SVR,t}^j, \bar{w}_{SVR,t-1}^j\right\}_{j=1,\cdots,M} . \tag{47}$$

**Theorem 1**: $\left\{\hat{\boldsymbol{x}}_t^i, w_{t-1}^i\right\}_{i=1,\cdots,N}$ and $\left\{\hat{\boldsymbol{x}}_{SVR,t}^j, \bar{w}_{SVR,t-1}^j\right\}_{j=1,\cdots,M}$ describe the same distribution in theory. (see Appendix A for a proof)

Although the new particles change the particle position distribution compared with that for the original particles, Theorem 1 shows that the PDF described by the new particles and their weights does not change compared with that of the original particles and their weights. The steps involved in region amplification and particle migration are summarized in Table I. After the region-amplification procedure and the particle migration technique, the even-distributed particles avoid the problem of over-dependence on the proposal distribution, thereby solving the degeneracy and impoverishment problem perfectly.

### *C. SVRPF ALGORITHM*

The analysis above shows that these new particles and their weights may be used to describe the prior-PDF. To describe the posterior-PDF, likelihood weights of these new particles must first be calculated,

$$\tilde{w}_{SVR,t}^j = \frac{p\left(\boldsymbol{y}_t \middle| \hat{\boldsymbol{x}}_{SVR,t}^j\right)}{\sum_{j=1}^{M} p\left(\boldsymbol{y}_t \middle| \hat{\boldsymbol{x}}_{SVR,t}^j\right)}, \tag{48}$$

and then the mixed weights from

$$\hat{w}_{SVR,t}^j = \frac{\tilde{w}_{SVR,t}^j \bar{w}_{SVR,t-1}^j}{\sum_{j=1}^{N} \tilde{w}_{SVR,t}^j \bar{w}_{SVR,t-1}^j} . \tag{49}$$

The state is estimated from

$$\bar{\boldsymbol{x}}_{SVR,t} = \sum_{j=1}^{M} \hat{w}_{SVR,t}^j \hat{\boldsymbol{x}}_{SVR,t}^j . \tag{50}$$

Regarding the integral structures of SVRPF and GPF (Fig. 3), the positioning of the particles in the GPF depends on the prior-PDF completely. When the likelihood-PDF is narrow and far from the high probability of the prior-PDF, particles near the high probability of the likelihood-PDF are very sparse. In this instance, details of the likelihood distribution may be lost. SVRPF estimates the prior-PDF by using SVR, and uniformly places particles in SVR-IR before receiving the current observation, which greatly reduces the risk of information loss of the likelihood distribution. Meanwhile, compared with EPF and UPF, the current observation is not considered in this procedure. Hence, the timeliness of SVRPF is not greatly affected.

**TABLE II**
SVRPF ALGORITHM

Initialization: $w_0^i \leftarrow N^{-1}$, $\boldsymbol{x}_0^i \leftarrow p(\boldsymbol{x}_0)$

// Over all time steps:

**for** $t \leftarrow 1$ to $T$ do

// Over all particles:

**for** $i \leftarrow 1$ to $N$ do

1. predict particles $\hat{\boldsymbol{x}}_t^i$ that satisfy the prior distribution according to (1)

**End**

2. perform region amplification and particle migration according to Table I
3. evaluate likelihood weights according to (48)
4. calculate weights commixture and normalization according to (49)
5. output the estimates from (50)
6. resample $N$ particles from the $M$ new particles according to the mixed weights and reset weights to $w_t^i = N^{-1}$.

**End**

Details of the SVRPF steps are summarized in Table II.

**Theorem 2:** Suppose the distribution function supported by $\left\{\hat{\boldsymbol{x}}_{SVR,t}^j, \tilde{w}_{SVR,t}^j\right\}$ is $p_{Li}(\boldsymbol{x}_t)$, then $p_{Li}(\boldsymbol{x}_t) \approx p(\boldsymbol{x}_t|\boldsymbol{y}_{1:t})$ (A proof is given in Appendix A.)

**Theorem 3:** Suppose the distribution function supported by $\left\{\hat{\boldsymbol{x}}_{SVR,t}^j, \hat{w}_{SVR,t}^j\right\}$ is $p_{Pos}(\boldsymbol{x}_t)$ then $p_{Pos}(\boldsymbol{x}_t) \approx p(\boldsymbol{x}_t|\boldsymbol{y}_{1:t})$ (A proof is given in Appendix A.)

In regard to Theorems 1–3, and the consistency of the resampling technique [46], SVRPF is seen in theory to be consistent with Bayes filtering. The essence of the SVRPF is that it generates uniformly-dispersed particles in the SVR-***IR*** through the particle migration technique, thereby reducing the risk of information loss regarding the likelihood distribution. Nevertheless, by employing the region-amplification technique, the SVRFP describes the posterior-PDF with greater stability.

## V. NUMERICAL SIMULATION

To assess the performance of the SVRPF for the nonlinear and non-Gaussian dynamic models, a classic dynamic model is used in this experiment [21]

$$x_t = 1 + \sin(0.04\pi t) + 0.5 x_{t-1} + u_t, \tag{51}$$

$$y_t = 0.2 x_t^2 + v_k, \tag{52}$$

where $u_t \sim \boldsymbol{\Gamma}(3,2)$, $v_t \sim \boldsymbol{N}(0, \sigma_v^2)$, and $x_0^i \sim \boldsymbol{N}(0.1, \sqrt{2})$.

Here, both the system and the measurement equation are strongly nonlinear. To study further the effect of observation noise on estimation performance, the experiment was conducted with two settings $\sigma_v^2 = 0.1$ *and* $5$.

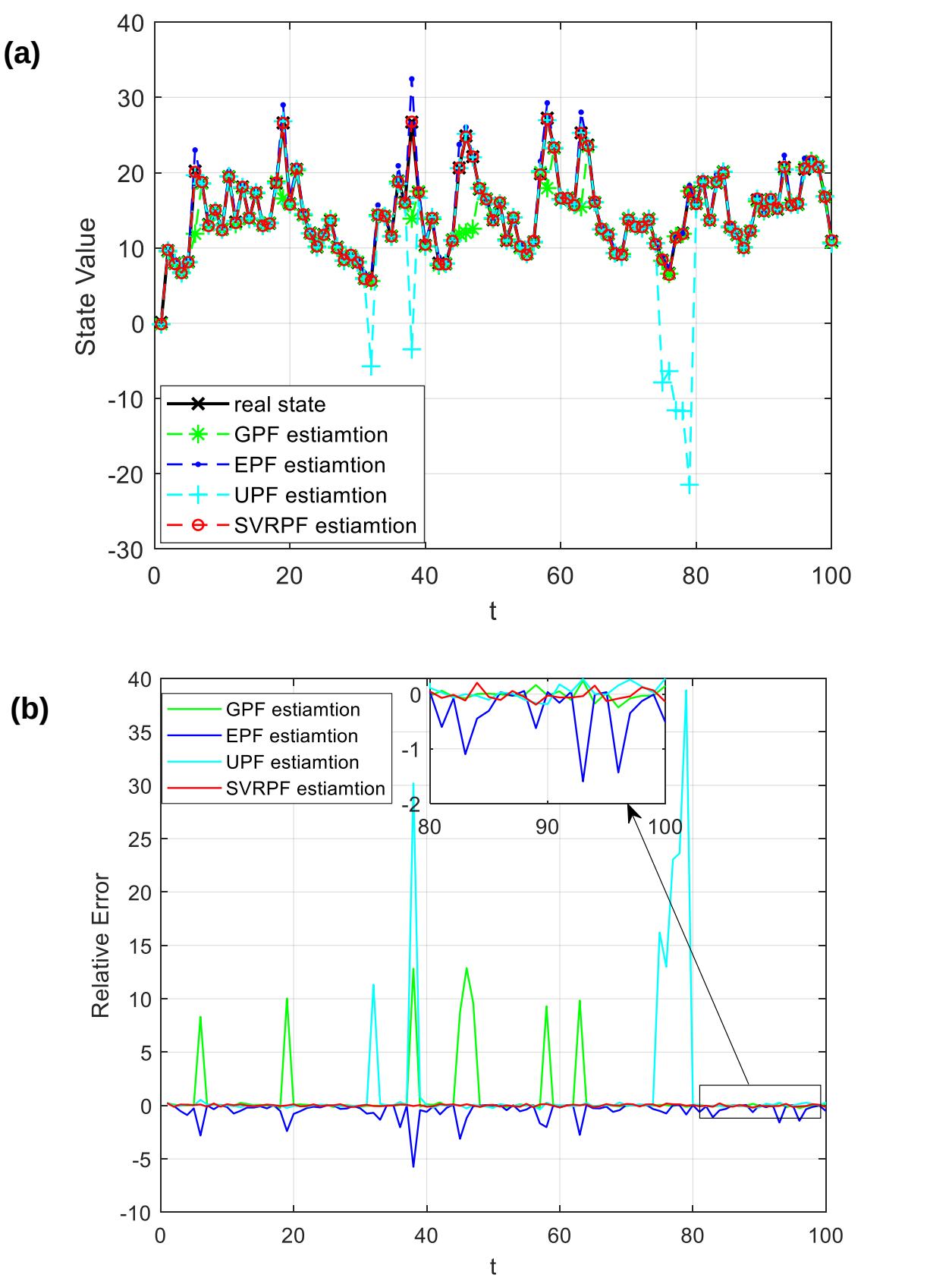

**FIGURE 4.** **Stability analysis of different particle filters in a narrow observation noise environment: (a) state estimation results and (b) estimated relative errors.**

To evaluate the stability of the filtering algorithms, we use the root-mean-square error (RMSE)

$$RMSE = \sqrt{\frac{1}{T}\sum_{t=1}^{T}\left[x_{real}(t) - x_{estimation}(t)\right]^2}, \tag{53}$$

where $T$ represents the simulation time step ($T$=100); the number of Monte Carlo runs is 100. In the SVRPF, $\eta = 0.5$ is used in region amplification and particle migration (Table I) and, in fitting the prior-PDF $\lambda = 1.8$, is used in the kernel function. The number of set particles is always equal to the original number of particles, that is, *M*=*N*.

In the next two experiments, we compare the performance of SVRPF to that of GPF [4,], EPF [18], and UPF [19] in a single simulation experiment, and to that of GPF, minimum variance resampling PF [32], residual resampling PF [29], EPF, UPF, branching PF [34] in the Monte Carlo simulation experiment.

### *A. Narrow Observation Noise with* $\sigma_v^2$=0.1

For the single simulation experiment, the number of particles is 100. From the results of the state estimation and the relative error (Fig. 4) for each algorithm, the UPF is the most affected by the strong nonlinearity of the observation, although the GPF has a large estimation deviation in some time steps. The frequency of the large deviation for the EPF is much higher than that of the GPF, although the amplitude of this deviation is smaller. The SVRPF has the more stable performance and the smallest relative error.

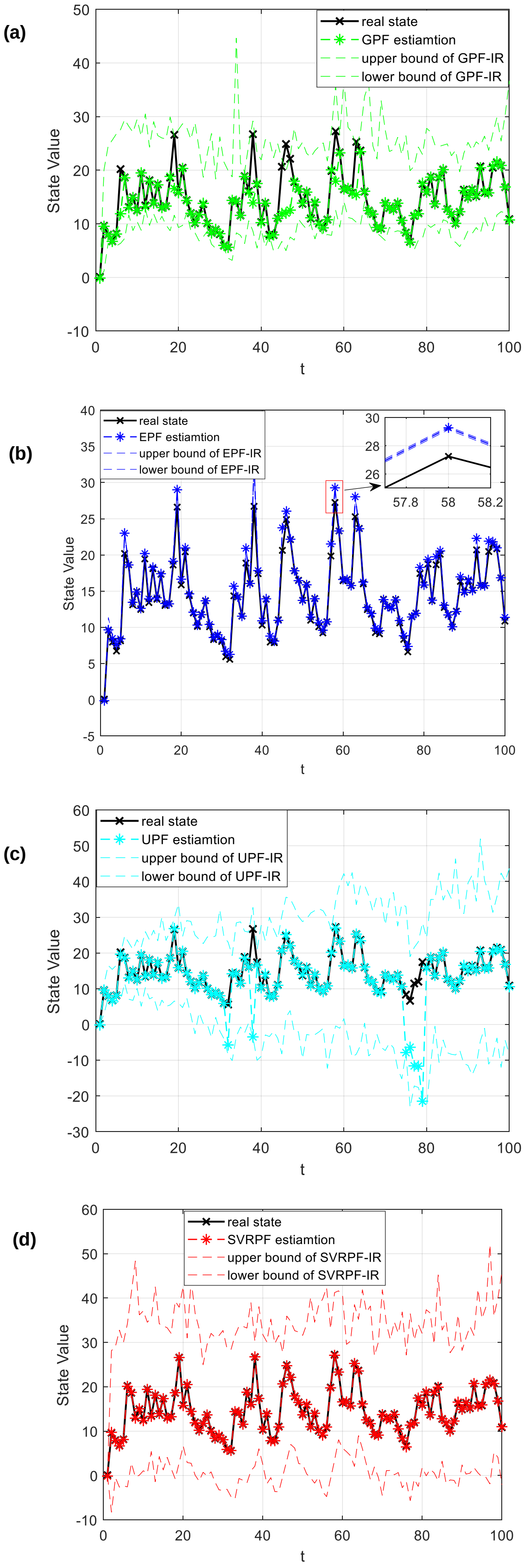

**FIGURE 5.** ***IR* for an environment with narrow observation noise for the different PFs: (a) GPF-*IR* (b) EPF-*IR* (c) UPF-*IR* (d) SVRPF-*IR*.**

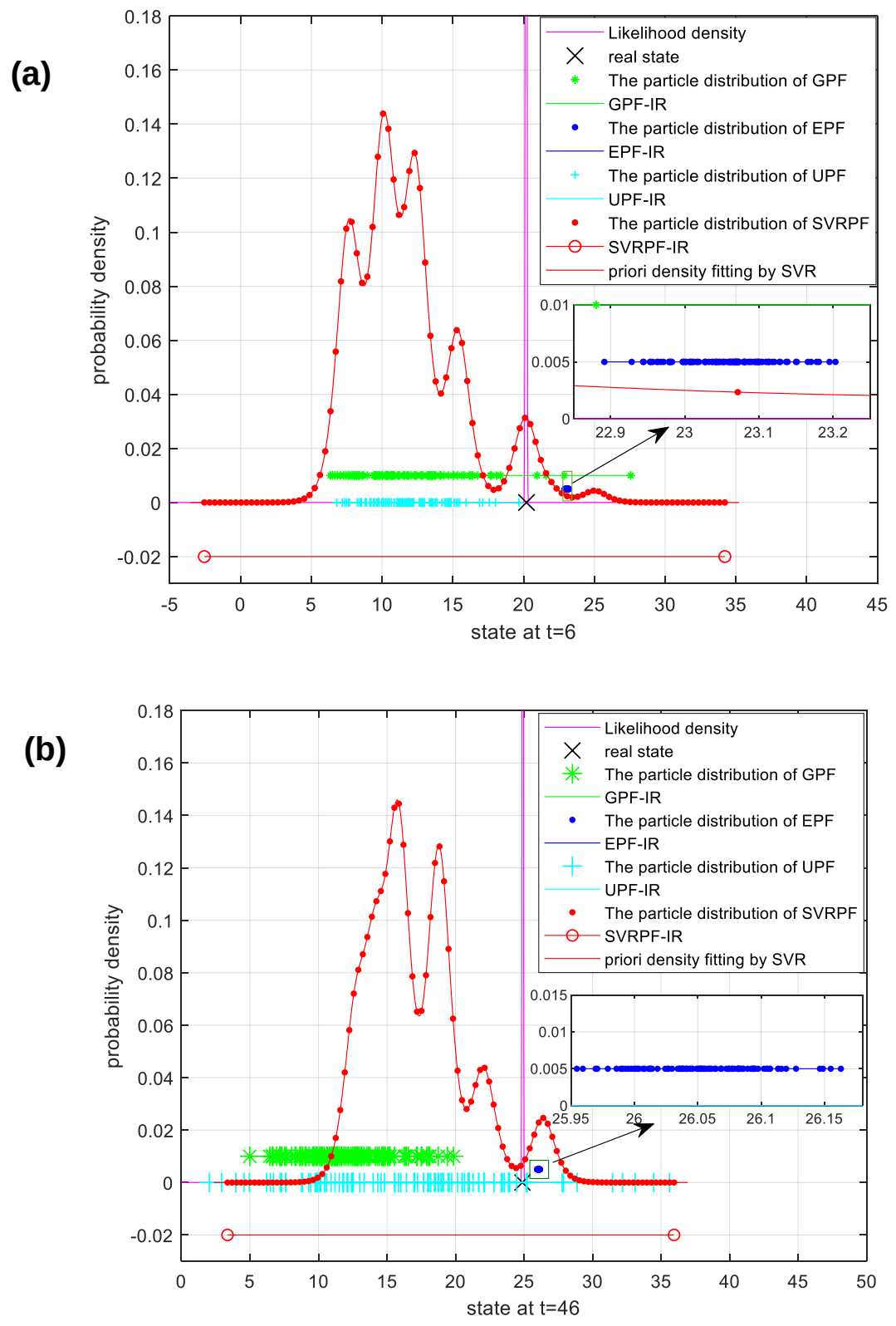

**FIGURE 6. Distribution of particles and its description of the PDF at time steps (a) *t*=6 and (b) *t*=46.**

To analyze further the performance of the different algorithms, the ***IR***s of the filtering algorithms and the distribution of particles for some time steps were plotted (Figs. 3 and 4, respectively).

From Fig. 5, the real state can be contained in the GPF-***IR*** in most instances, but not always. The EPF-***IR*** is too narrow that it cannot contain the real state in most instances, although it is close to the real state. The UPF-***IR*** grows with increasing time steps, and the symmetry of the observation equation interferes strongly with the performance of the UPF. The SVR-***IR***, which stably contains the real state at all times, comes from amplifying the GPF-***IR***.

From the distribution of particles at time steps *t*=6 and *t*=46 [Fig. 6(a) and (b), respectively], the GPF-IR contains the real state at *t*=6 but fails at *t*=46. However, there are almost no particles in the region with a high likelihood-PDF at either *t*=6 or 46. This phenomenon weakens the effect of observation on state estimation, leading to a large deviation. The EPF concentrates the particles more in a small region. As the observation noise is small, the truncation error is particularly prominent and affects the accuracy of the likelihood weight. There is no particle in the high probability of the likelihood-PDF for the UPF, although the

UPF has a wide ***IR***.

Particles generated by the SVRPF are uniformly distributed in the SVR-***IR***. In this way, particles always exist in the high probability of the likelihood-PDF, even if the likelihood probability is very narrow and far from the high probability of the prior-PDF. This is the main reason for the stability of the SVRPF algorithm.

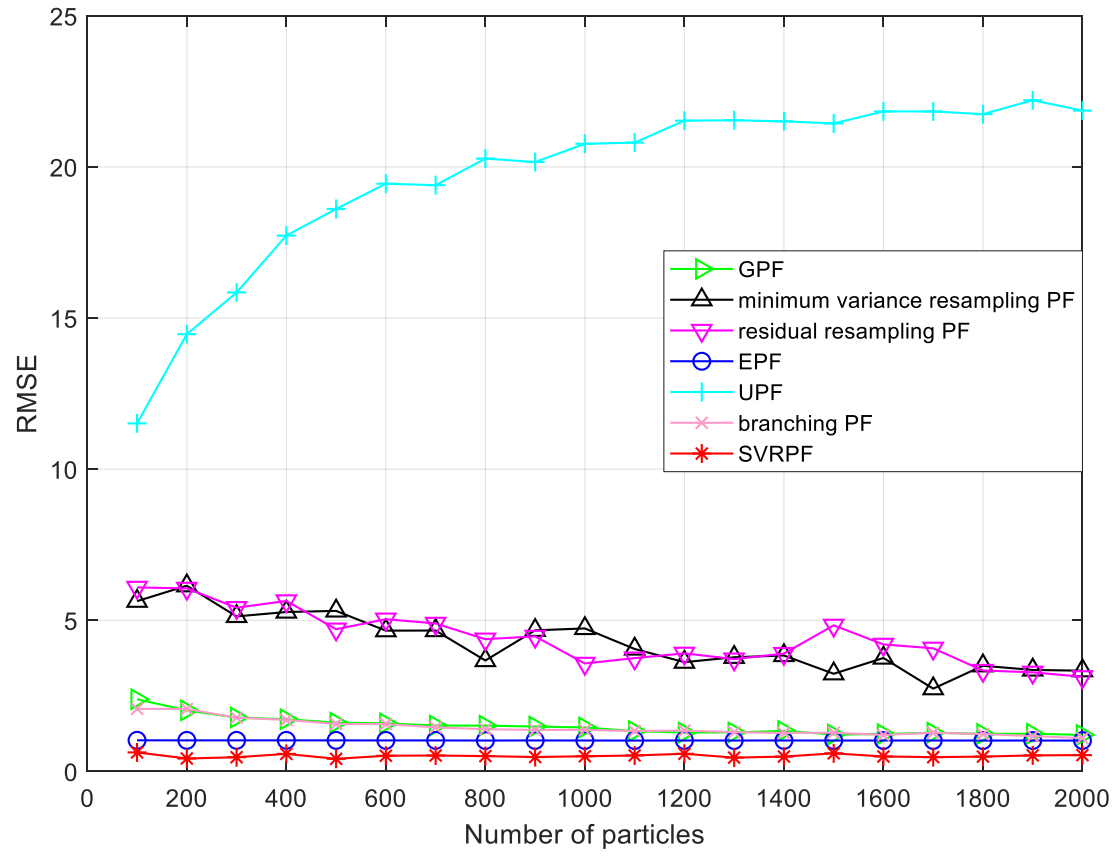

**FIGURE 7. Variation of RMSE with particle number for various PFs in an environment with narrow observation noise.**

Fig. 7 shows the RMSE of different algorithms with increasing particle number in an environment with narrow observation noise. For the branching PF, the number of particles means the initial number of particles. The performance of the UPF is the worst in consequence of the symmetry of the observation equation. The RMSE for the GPF is closer to the EPF as the number of particles increases. The RMSE for the GPF and branching PF is similar. The performances of the minimum variance resampling PF, and residual resampling PF are slightly worse compared with the GPF, branching PF, EPF, and SVRPF. The RMSE of the SVRPF in this instance is better than that of all other algorithms.

### *B. Normal Observation Noise with* $\sigma_v^2=5$

We see from Fig. 8 that all these filtering algorithms have similar performances except the UPF. Fig. 9 shows the SVR-***IR*** is very stable, and the real state is contained within it. The GPF-***IR*** is larger than that of the EPF whereas the UPF-IR grows with increasing time step.

However, the ***IR*** is just one of the elements that reflects the performance of the state estimation; another is the PD. Fig. 10 shows that the EPF-***IR*** at $t=89/93$ does not contain the real state because of an over-concentration of the particles and truncation errors. Although the particle distributions of the GPF and UPF are sufficiently dispersed, the non-uniform particle distribution might affect the stability of the state estimation. This instability is concentrated at moments when the proposal distribution is far from the real posterior-PDF. The SVR-***IR*** always contains the real state perfectly. Moreover, uniform particles help avoid performance instabilities caused by deviations in the proposed distribution.

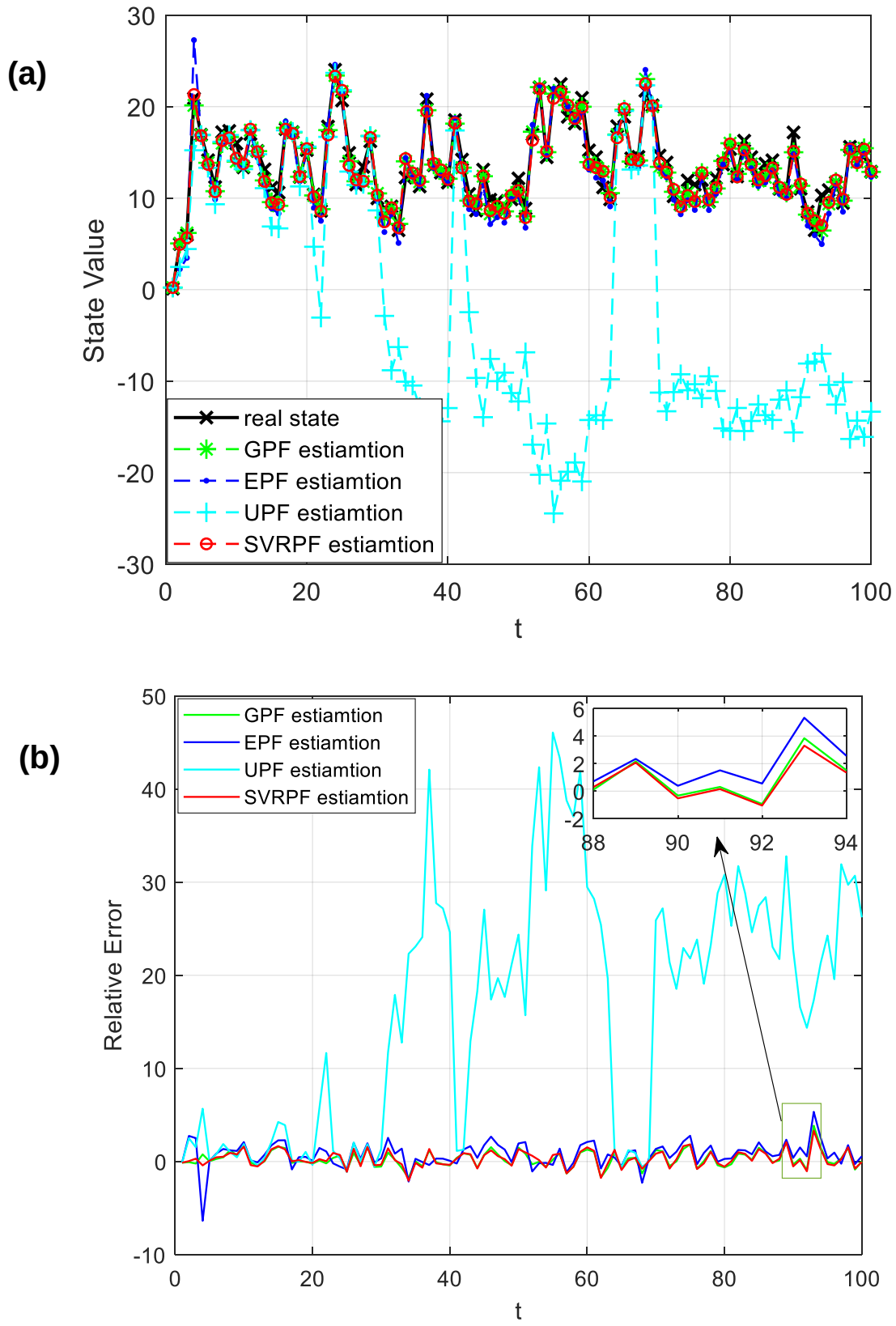

**FIGURE 8. Results generated by different particle filters for an environment with normal observation noise: (a) state estimation and (b) estimated relative errors.**

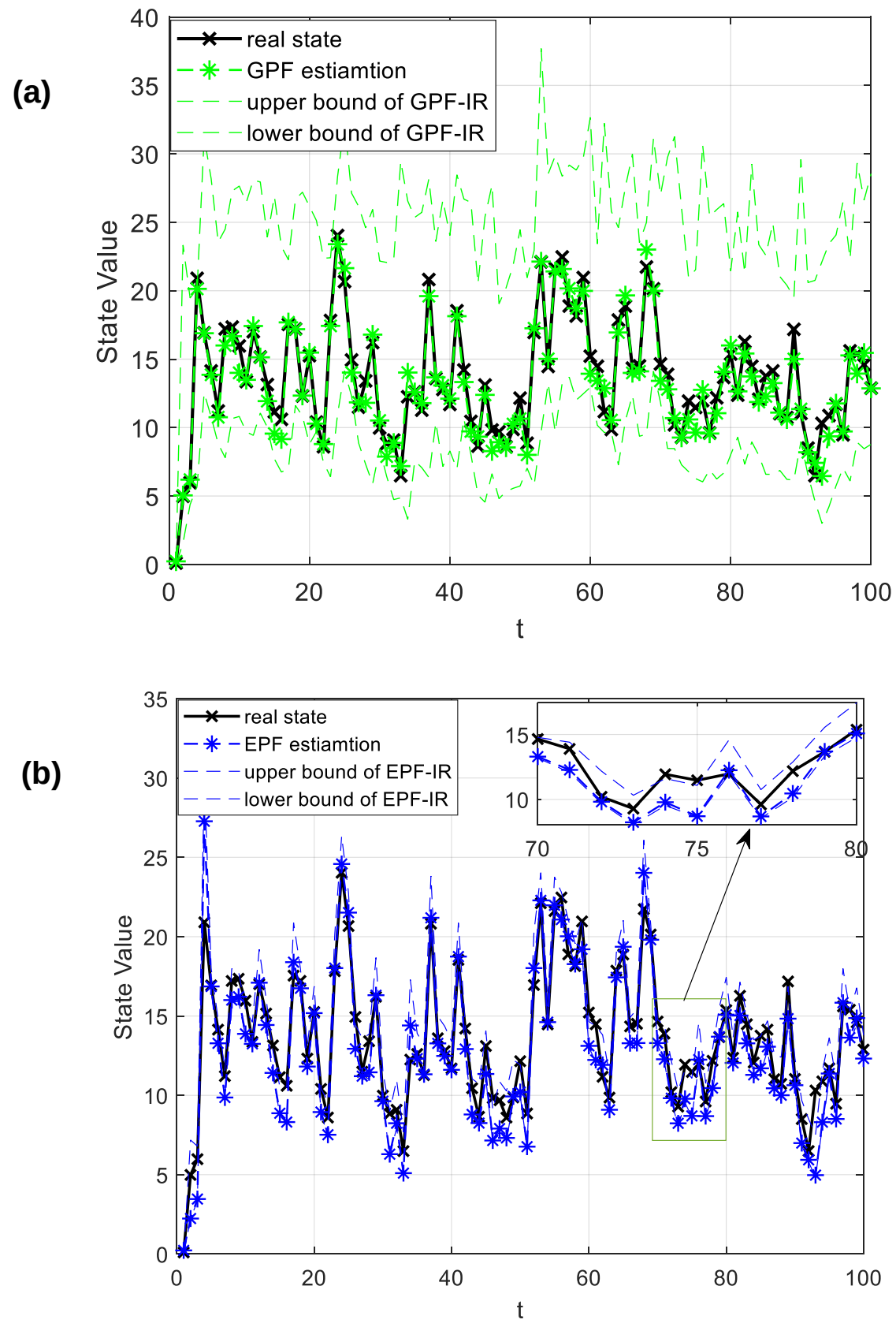

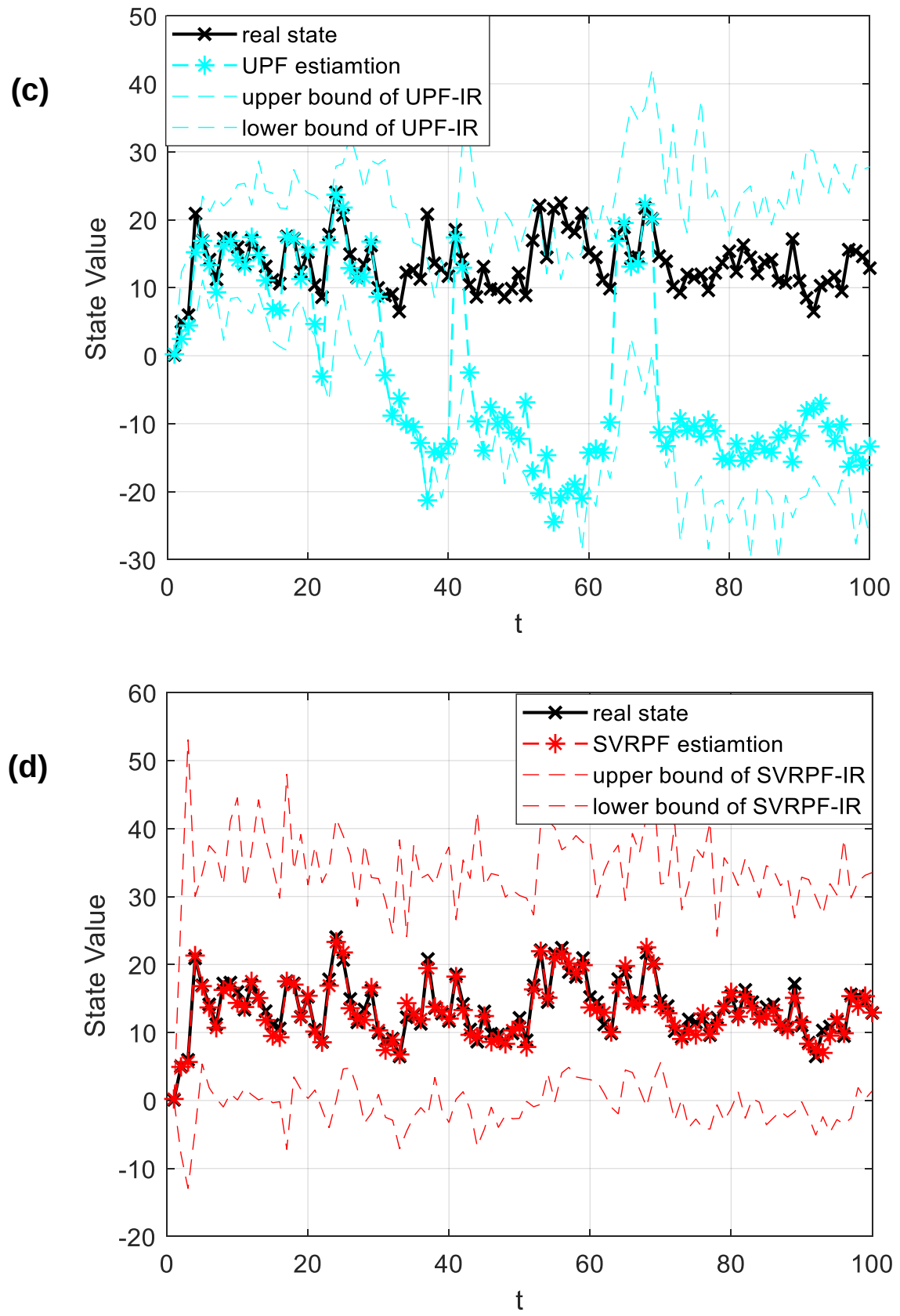

**FIGURE 9.** ***IR*s for an environment with normal observation noise for different particle filters: (a) GPF-*IR* (b) EPF-*IR* (c) UPF-*IR* (d) SVRPF-*IR*.**

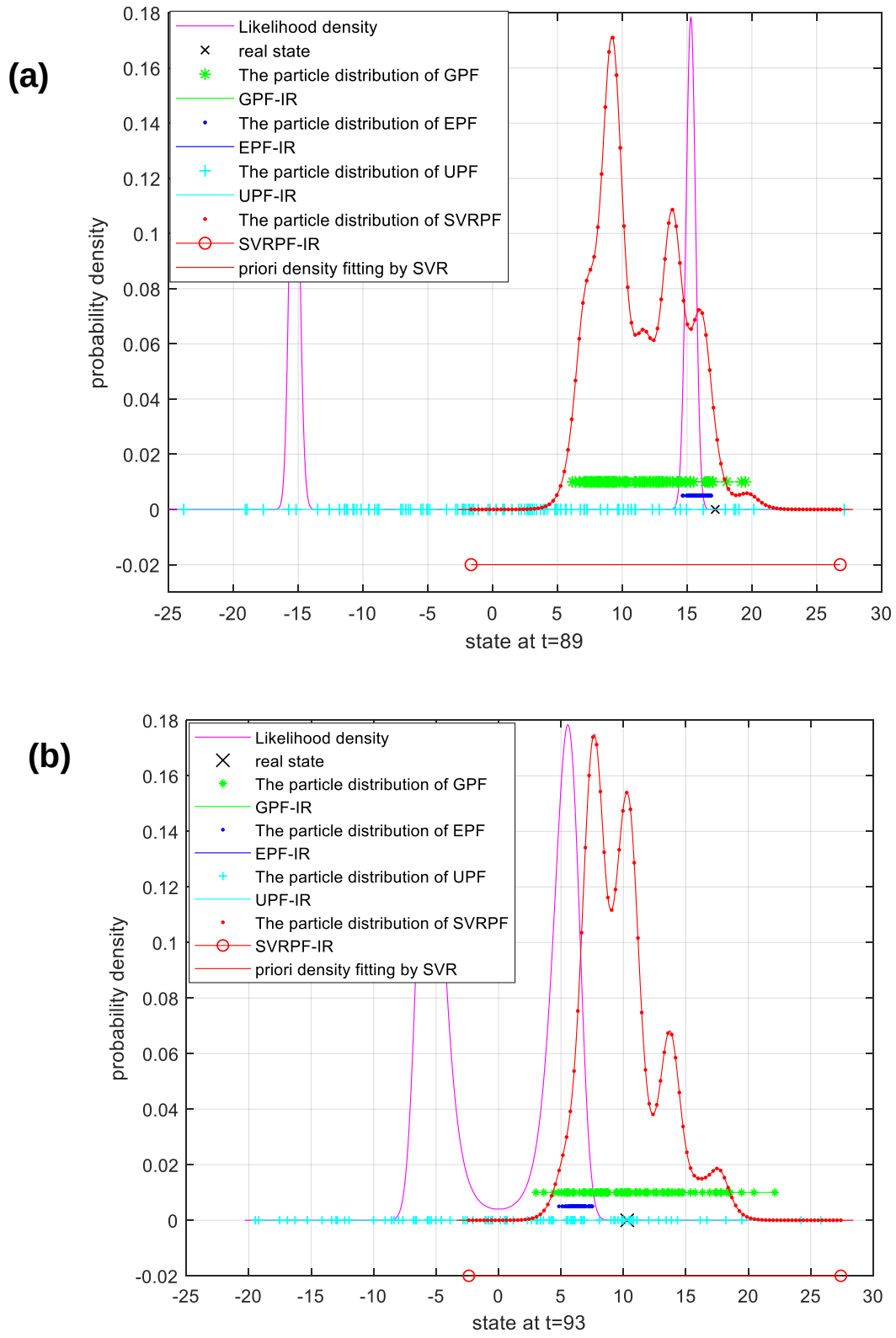

**FIGURE 10. Distribution of particles and its description of the PDF with normal observation noise: (a) t=89; (b) t=93**

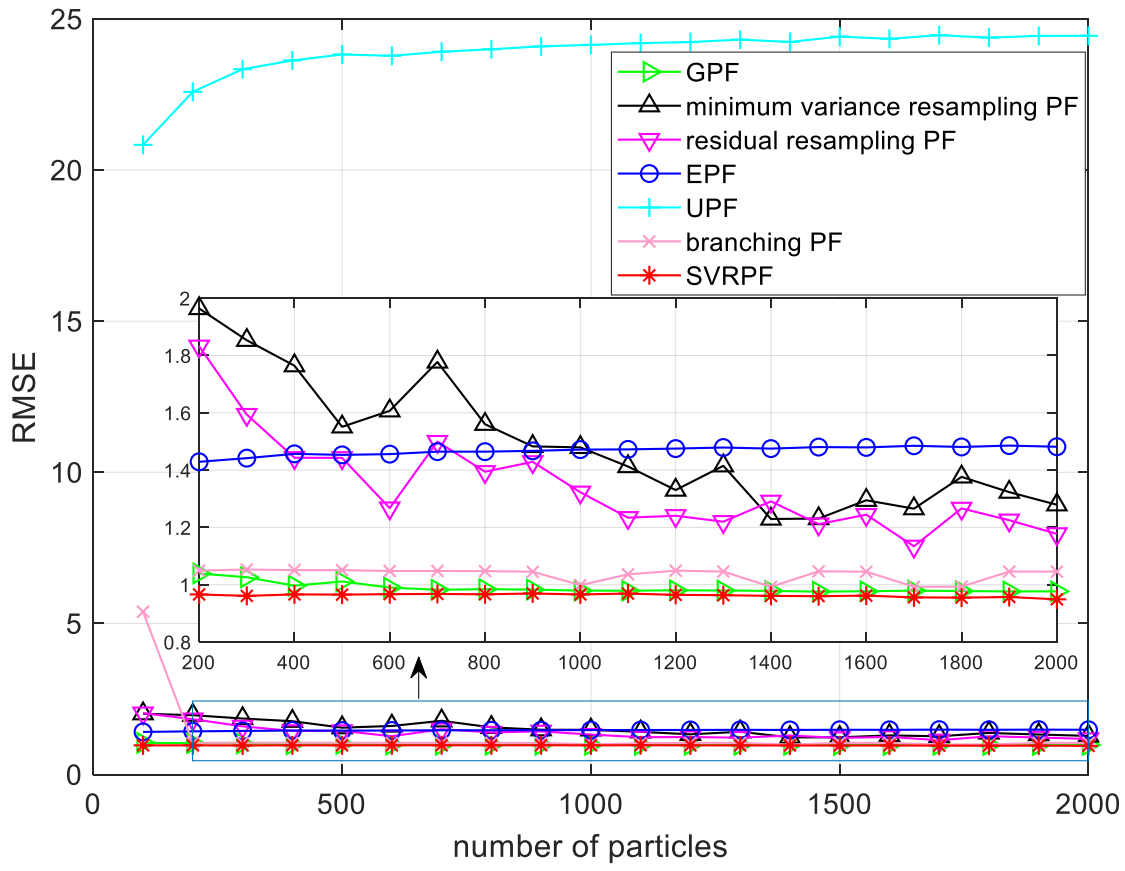

**FIGURE 11. Variation of RMSE in normal observation noise with particle number.**

The variation of the RMSE for different PFs in an environment with normal observation noise was plotted against particles number (Fig. 11). In contrast to the previous experiment, the performance of the GPF is better than that of the EPF in this instance. This is because large deviations disappear for the GPF compared with the previous experiment. Moreover, the performance of the UPF is still the worst in consequence of the divergent phenomenon with the UPF-***IR*** and the symmetry of the observation equation. The performance of the branching PF is also similar to that of the GPF. The performance of the minimum variance resampling PF and residual resampling PF are slightly worse, compared with the GPF, branching PF, and SVRPF. Finally, the performance of the SVRPF is better than that of other algorithms.

## V. CONCLUSION

Various PF algorithms were reanalyzed from a new perspective. The factors influencing their performance fall into two categories: ***IR*** selection and PD. Furthermore, the SVRPF was proposed to overcome difficulties from nonlinear and non-Gaussian environments, especially for narrow observation noise. The core of this method is to leverage the SVR to fit the expression of the prior-PDF, to disperse particles to ensure a uniform distribution in the SVR-***IR***, and to calculate the SVR weights of these new particles. The likelihood weights of these new particles can then be calculated and the state estimation obtained using these new particles and the mixed weights. In addition, the consistency of this algorithm and Bayes' theorem have been proved. Simulation experiments demonstrated that the performance of the proposed algorithm is perfect in nonlinear and non-Gaussian environments, especially those with narrow observation noise, although the computational time has increased.

## APPENDIX A

***Proof of Theorem 1:*** With the prior-PDF represented by $\left\{\hat{\boldsymbol{x}}_t^i, \bar{w}_{t-1}^i\right\}$, we have

$$F\left(\boldsymbol{x}_t \,|\boldsymbol{y}_{1:t-1}\right) = \lim_{N\to\infty}\sum_{i=1}^{N}\bar{w}_{t-1}^i\theta\left(\boldsymbol{x}_t \succ \hat{\boldsymbol{x}}_t^i\right), \tag{54}$$

where $F\left(\boldsymbol{x}_t \,|\boldsymbol{y}_{1:t-1}\right)$ represents the CDF of the prior-PDF. The CDF estimated by the SVR algorithm is found to be

$$F_{SVR}\left(\boldsymbol{x}_t\right) = \int_{-\infty}^{\boldsymbol{x}_t} p_{SVR}\left(\boldsymbol{u},\boldsymbol{\beta}_t\right)d\boldsymbol{u} \tag{55}$$

$$= \sum_{i=1}^{N}\beta_{t,i}K\left(\boldsymbol{x}_t, \hat{\boldsymbol{x}}_t^i\right) \tag{56}$$

$$= \sum_{i=1}^{N}\bar{w}_{t-1}^i\theta\left(\boldsymbol{x}_t \succ \hat{\boldsymbol{x}}_t^i\right) \tag{57}$$

$$\approx F\left(\boldsymbol{x}_t \,|\boldsymbol{y}_{1:t-1}\right) \tag{58}$$

Therefore, the distribution function $F_{SVR}\left(\boldsymbol{x}_t\right)$, which is obtained by the SVR probability density estimation, is the same as the distribution function $F\left(\boldsymbol{x}_t \,|\boldsymbol{y}_{1:t-1}\right)$.

The distribution function represented by these new particles and their weights is

$$F_{NP}\left(\boldsymbol{x}_t\right) = \sum_{j=1}^{M}\bar{w}_{SVR,t-1}^j\theta\left(\boldsymbol{x}_t \succ \hat{\boldsymbol{x}}_{SVR,t}^j\right) \tag{59}$$

$$= \frac{\sum_{j=1}^{M}\sum_{i=1}^{N}\beta_{t,i}k\left(\hat{\boldsymbol{x}}_{SVR,t}^j, \hat{\boldsymbol{x}}_t^i\right)\theta\left(\boldsymbol{x}_t \succ \hat{\boldsymbol{x}}_{SVR,t}^j\right)}{\sum_{j=1}^{M}\sum_{i=1}^{N}\beta_{t,i}k\left(\hat{\boldsymbol{x}}_{SVR,t}^j, \hat{\boldsymbol{x}}_t^i\right)} \tag{60}$$

$$= \frac{\sum_{j=1}^{M}p_{SVR}\left(\hat{\boldsymbol{x}}_{SVR,t}^j,\boldsymbol{\beta}_t\right)\theta\left(\boldsymbol{x}_t \succ \hat{\boldsymbol{x}}_{SVR,t}^j\right)}{\sum_{j=1}^{M}p_{SVR}\left(\hat{\boldsymbol{x}}_{SVR,t}^j,\boldsymbol{\beta}_t\right)} \tag{61}$$

In (61), the numerator represents the sum of the probability densities of particles that satisfy $\theta\left(\boldsymbol{x}_t \succ \hat{\boldsymbol{x}}_{SVR,t}^j\right)$; the denominator is the normalization coefficient. As new particles are homogeneous in SVR-***IR***, (61) becomes

$$\frac{\sum_{j=1}^{M}p_{SVR}\left(\hat{\boldsymbol{x}}_{SVR,t}^j,\boldsymbol{\beta}_t\right)\theta\left(\boldsymbol{x}_t \succ \hat{\boldsymbol{x}}_{SVR,t}^j\right)}{\sum_{j=1}^{M}p_{SVR}\left(\hat{\boldsymbol{x}}_{SVR,t}^j,\boldsymbol{\beta}_t\right)} \approx \int_{-\infty}^{\boldsymbol{x}_t} p_{SVR}\left(\boldsymbol{u},\boldsymbol{\beta}_t\right)d\boldsymbol{u} \tag{62}$$

$$= F_{SVR}\left(\boldsymbol{x}_t\right). \tag{63}$$

We then have

$$F\left(\boldsymbol{x}_t \,|\boldsymbol{y}_{1:t-1}\right) \approx F_{SVR}\left(\boldsymbol{x}_t\right) \approx F_{NP}\left(\boldsymbol{x}_t\right). \tag{64}$$

Hence, the distribution of the new particles and their weights are the same as that of the original particles and their weights.

***Proof of Theorem 2:*** Similar to Theorem 1, the CDF of the distribution that was described by $\left\{\hat{\boldsymbol{x}}_{SVR,t}^j, \tilde{w}_{SVR,t}^j\right\}$ is written

$$F_{Li}\left(\boldsymbol{x}_t\right) = \sum_{j=1}^{M}\tilde{w}_{SVR,t}^j\theta\left(\boldsymbol{x}_t \succ \hat{\boldsymbol{x}}_{SVR,t}^j\right) \tag{65}$$

$$= \frac{\sum_{j=1}^{M}p\left(\boldsymbol{y}_t \,|\hat{\boldsymbol{x}}_{SVR,t}^j\right)\theta\left(\boldsymbol{x}_t \succ \hat{\boldsymbol{x}}_{SVR,t}^j\right)}{\sum_{j=1}^{M}p\left(\boldsymbol{y}_t \,|\hat{\boldsymbol{x}}_{SVR,t}^j\right)}. \tag{66}$$

As new particles are homogeneous distributed in SVR-***IR***, (66) can be rewritten as

$$\frac{\sum_{j=1}^{M}p\left(\boldsymbol{y}_t \,|\hat{\boldsymbol{x}}_{SVR,t}^j\right)\theta\left(\boldsymbol{x}_t \succ \hat{\boldsymbol{x}}_{SVR,t}^j\right)}{\sum_{j=1}^{M}p\left(\boldsymbol{y}_t \,|\hat{\boldsymbol{x}}_{SVR,t}^j\right)} \approx \int_{-\infty}^{\boldsymbol{x}_t} p\left(\boldsymbol{y}_t \,|\boldsymbol{u}\right)d\boldsymbol{u} \tag{67}$$

$$= F\left(\boldsymbol{y}_t \,|\boldsymbol{x}_t\right). \tag{68}$$

Hence, $p_{Li}\left(\boldsymbol{x}_t\right) \approx p\left(\boldsymbol{y}_t \,|\boldsymbol{x}_t\right)$. □

***Proof of Theorem 3:*** The CDF of the distribution described by $\left\{\hat{\boldsymbol{x}}_{SVR,t}^j, \hat{w}_{SVR,t}^j\right\}$ is written

$$F_{Pos}\left(\boldsymbol{x}_t\right) = \sum_{j=1}^{M}\hat{w}_{SVR,t-1}^j\theta\left(\boldsymbol{x}_t \succ \hat{\boldsymbol{x}}_{SVR,t}^j\right), \tag{69}$$

in accordance with (46), (48), and (49); (69) can be rearranged to give

$$= \frac{\sum_{j=1}^{M}\tilde{w}_{SVR,t}^j\bar{w}_{SVR,t-1}^j\theta\left(\boldsymbol{x}_t \succ \hat{\boldsymbol{x}}_{SVR,t}^j\right)}{\sum_{j=1}^{M}\bar{w}_{SVR,t-1}^j\hat{w}_{SVR,t}^j}, \tag{70}$$

$$= \frac{\sum_{j=1}^{M}p\left(\boldsymbol{y}_t \,|\hat{\boldsymbol{x}}_{SVR,t}^j\right)p_{SVR}\left(\hat{\boldsymbol{x}}_{SVR,t}^j,\boldsymbol{\beta}_t\right)\theta\left(\boldsymbol{x}_t \succ \hat{\boldsymbol{x}}_{SVR,t}^j\right)}{\sum_{j=1}^{M}p_{SVR}\left(\hat{\boldsymbol{x}}_{SVR,t}^j,\boldsymbol{\beta}_t\right)p\left(\boldsymbol{y}_t \,|\hat{\boldsymbol{x}}_{SVR,t}^j\right)}, \tag{71}$$

as new particles are distributed homogeneously in SVR-IR. Rearranging (71) yields

$$\frac{\sum_{j=1}^{M}p\left(\boldsymbol{y}_t \,|\hat{\boldsymbol{x}}_{SVR,t}^j\right)p_{SVR}\left(\hat{\boldsymbol{x}}_{SVR,t}^j,\boldsymbol{\beta}_t\right)\theta\left(\boldsymbol{x}_t \succ \hat{\boldsymbol{x}}_{SVR,t}^j\right)}{\sum_{j=1}^{M}p_{SVR}\left(\hat{\boldsymbol{x}}_{SVR,t}^j,\boldsymbol{\beta}_t\right)p\left(\boldsymbol{y}_t \,|\hat{\boldsymbol{x}}_{SVR,t}^j\right)} \tag{72}$$

$$\approx \int_{\boldsymbol{x}_{\min}}^{\boldsymbol{x}_t} p\left(\boldsymbol{x}_t \,|\boldsymbol{y}_{1:t}\right)d\boldsymbol{u} = F\left(\boldsymbol{x}_t \,|\boldsymbol{y}_{1:t}\right) \tag{73}$$

Hence, $p_{Pos}\left(\boldsymbol{x}_t\right) \approx p\left(\boldsymbol{x}_t \,|\boldsymbol{y}_{1:t}\right)$. □